\documentclass[aps,pre,twocolumn,superscriptaddress,preprintnumbers,amsmath,amssymb,floatfix]{revtex4-1}

\usepackage{grffile}
\usepackage{graphicx}
\usepackage{stackrel}

\newlength{\onefig}
\newlength{\twofig}

\usepackage[usenames,dvipsnames]{color}

\begin{document}

\title{Cluster and reentrant anomalies of nearly Gaussian core particles}

\author{Daniele Coslovich}
\email{Email: daniele.coslovich@univ-montp2.fr}
\thanks{corresponding author}
\affiliation{Universit{\'e} Montpellier 2, Laboratoire Charles Coulomb UMR 5221, Montpellier, France}
\affiliation{CNRS, Laboratoire Charles Coulomb UMR 5221, Montpellier, France}

\author{Atsushi Ikeda}
\affiliation{Universit{\'e} Montpellier 2, Laboratoire Charles Coulomb UMR 5221, Montpellier, France}
\affiliation{CNRS, Laboratoire Charles Coulomb UMR 5221, Montpellier, France}

\date{\today}

\begin{abstract}
We study through integral equation theory  and  numerical simulations the structure and dynamics of fluids composed of ultrasoft, nearly Gaussian particles.
Namely, we explore the fluid phase diagram of a model in which particles interact via the generalized exponential potential $u(r)=\epsilon \exp[-(r/\sigma)^n]$, with a softness exponent $n$ slightly larger than 2.
In addition to the well-known anomaly associated to reentrant melting, the structure and dynamics of the fluid display two additional anomalies, which are visible in the isothermal variation of the structure factor and diffusivity.
These features are correlated to the appearance of dimers in the fluid phase and to the subsequent modification of the cluster structure upon compression.
We corroborate these results through an analysis of the local minima of the potential energy surface, in which clusters appear as much tighter conglomerates of particles.
We find that reentrant melting and clustering coexist for softness exponents ranging from $2^+$ up to values relevant for the description of amphiphilic dendrimers, i.e., $n=3$.
.
\end{abstract}


\maketitle

\section{Introduction}\label{sec:intro}

Hard colloidal suspensions have been extensively studied as the simplest model of particulate systems.
In contrast, several colloidal suspensions are composed of soft, macroscopic particles, 
and display a much more complex behavior~\cite{Likos_2006, Vlassopoulos_Cloitre_2012}.
Star polymers~\cite{Likos1998,zaccarelli_tailoring_2005,stiakakis_slow_2010} and microgels~\cite{Saunders_Vincent_1999,Senff_Richtering_1999,Lietor-Santos_Sierra-Martin_Vavrin_Hu_Gasser_Fernandez-Nieves_2009} are notable examples of purely repulsive, soft colloids, 
in which softness can be tuned by either altering the macromolecular architecture or by exploiting the sensitivity to temperature and solvent properties. 
As a consequence these macromolecular aggregates can be compressed to attain very large packing fractions---significantly 
above random close packing~\cite{mattsson_soft_2009,Koumakis_Petekidis_2012}.

In theory and simulation, soft colloids can be studied by modeling them as point particles interacting through a soft repulsive potential $u(r)$. 
Paradigmatic examples are Hertzian particles~\cite{Pamies_Cacciuto_Frenkel_2009}, whose repulsion originates from the elastic deformation of spheres at contact, and the Gaussian core model (GCM)~\cite{Stillinger1976, Stillinger1979}.
The thermodynamic and dynamic behaviors of fluids composed of such soft particles differ significantly from the usual hard sphere model and display an anomalous, reentrant density dependence~\cite{Stillinger1976,Stillinger1979,louis_mean-field_2000,lang_fluid_2000,prestipino_phase_2005,Mausbach2006,mayer_asymmetric_2008,krekelberg_anomalous_2009,krekelberg_generalized_2009,Pamies_Cacciuto_Frenkel_2009,Jacquin_Berthier_2010,berthier_increasing_2010,ikeda_glass_2011,Ikeda_therm_2011,ikeda_slow_2011}.
Upon increasing density from the dilute regime, the particles first pack more densely and the structural order increases---as in hard spheres. 
Upon further compression, particles tend to decorrelate spatially, 
since the energetic gain of avoiding overlaps is overwhelmed by the increase of available configurations (entropic effect)~\cite{krekelberg_anomalous_2009,Jacquin_Berthier_2010}. 
This leads to looser spatial correlations and enhanced diffusivity.
From a physical point of view, the anomalous behavior can thus be understood in terms of the competition between entropic and energetic effects.
The anomalous behavior of the fluid has a direct counterpart in the thermodynamic phase diagram, which displays reentrant melting upon increasing density~\cite{prestipino_phase_2005,Pamies_Cacciuto_Frenkel_2009,Ikeda_therm_2011}.
Very recently, such anomalies have indeed been observed in experiments on microgels, which can be modeled rather well using Hertzian particles~\cite{Paloli_Mohanty_Crassous_Zaccarelli_Schurtenberger_2013}.

On the other hand, a different type of anomaly can occur if the pair potential is bounded at the origin, i.e., $u(r=0)$ is finite, but has a steep hill at some length scale $\sigma$.
An ideal example is the penetrable sphere model (PSM) in which $u(r)$ is a finite constant at $r < \sigma$ and zero otherwise~\cite{Marquest_1989}. 
At sufficiently high density, such ultrasoft particles display \textit{cluster formation}.
In the high-density, fluid phase, the radial distribution function shows a 
clear peak at $r=0$, meaning that the particles can interpenetrate and form relatively stable clusters~\cite{Likos_PSM_1998}. 
Depending on the details of the interaction potential, formation of cluster crystals~\cite{Likos_PSM_1998,mladek_formation_2006,mladek_clustering_2007,Zhang_Charbonneau_Mladek_2010,Zhang_PSM_2012} and cluster glasses~\cite{coslovich_cluster_2012} have also been observed at high density.
In such cluster phases, the collective structure is essentially frozen, but individual particles may diffuse by hopping on the cluster sites~\cite{moreno_diffusion_2007,coslovich_hopping_2011}. 
Through a mean-field analysis, Likos et al.~\cite{Likos_Lang_Watzlawek_Lowen_2001,likos_why_2007} have proposed that clustering occurs whenever the Fourier transform of potential, $u(k)$, has some negative components ($Q^{\pm}$ potentials). 

Both the reentrant anomalies and clustering can be studied within a coherent model system, called the generalized exponential model (GEM)
\begin{equation}\label{eqn:u}
u(r) = \epsilon \exp[-(r/\sigma)^n]
\end{equation}
where $n$ is a softness exponent~\cite{mladek_formation_2006}. 
Clearly, the GCM and the PSM are recovered when $n=2$ and $n=\infty$, respectively.
The GEM has been extensively studied for $n=4$ and $8$, but very little is known about the behavior for other values of $n$.
According to the mean-field argument of Likos et al.~\cite{Likos_Lang_Watzlawek_Lowen_2001,likos_why_2007}, clustering occurs for $n > 2$. 
Thus, the model allows to interpolate continuously between reentrant anomalies and cluster formation, with an interesting crossover expected around $n=2$. 
Moreover, values of $n$ around 3 are of interest, as they may be used to describe the effective interactions between macromolecules 
with low fractal dimensions, such as amphiphilic dendrimers~\cite{mladek_computer_2008}. 
Although these effective interactions are derived in the infinite dilute regime, a recent study indicates that their phenomenology may be observed in fully atomistic simulation of dendrimers at finite concentrations~\cite{Lenz_Cluster_2012}

In this paper we present a theoretical and numerical  investigation of the GEM-$n$ with varying softness $n \geq 2$. 
We show that the structural and dynamic properties display both reentrant and cluster anomalies when $n$ is larger than but close to 2. 
First, we conduct a preliminary investigation of the fluid phase diagram by means of integral equation theory for various values of $n$. 
We focus then on the behavior of the model with $n=2.2$. 
We find that the fluid structure and dynamics display a sequence of anomalies, which manifests themselves as multiple minima and maxima in the isothermal properties.
We find that the origin of the anomalies observed at high density are due to formation of clusters in the fluid. 
To provide a sharp characterization of cluster formation we analyze the potential energy surface of the model, 
following numerical techniques well-established in the context of supercooled liquids~\cite{sastry__1998}. 
An analysis of the local minima sampled above the melting temperature allows to obtain a coherent picture of the structural and dynamic anomalies visible at high density. 
 
\section{Methods}\label{sec:methods}

We consider a model of point particles interacting via the generalized exponential potential of order $n$, Eq.~\eqref{eqn:u}. As usual, $\epsilon$, $\sigma$, and $\sqrt{m\sigma^2/\epsilon}$ provide the units of length, energy, and time, respectively. We use integral equation theory to predict the static properties over a broad range of state parameters and $n$ values. The Ornstein-Zernike equation with the hypernetted chain closure (HNC) relation~\cite{hansen_theory_2006-1} was solved by an iterative algorithm. We also used the mean spherical approximation (MSA) closure relation~\cite{louis_mean-field_2000,likos_why_2007}, which can be solved analytically. Molecular dynamics simulations were carried for systems composed of $N=4000$ particles enclosed in a cubic box with periodic boundary conditions. Fluid samples were prepared starting from a random configuration using a Berendsen thermostat to adjust the temperature during the equilibration phase. At low temperature, additional runs were carried out using a slow annealing to make sure that static and dynamic properties were reproducible and were not affected by partial crystallization or phase separation. State points for which at least one sample crystallized were not considered in our analysis, which only focused on the fluid portion of the phase diagram. Production runs were performed in the microcanonical ensemble using the velocity-Verlet algorithm. The integration time step was 0.05 in reduced units for both equilibration and production runs. The properties of the GCM are known to be sensitive to the cut off radius $r_c$ of the potential~\cite{ikeda_slow_2011,ingebrigtsen_what_2012}. Therefore, a relatively long range cut off was applied ($r_c=4.0$). In addition, we employed a cubic spline interpolation~\cite{grigera_geometric_2002-1} between $r_c=4.0$ to $4.17$ to improve energy conservation during the MD simulations and to avoid numerical difficulties in the analysis of the potential energy surface. The local minima (or inherent structures~\cite{stillinger__1982}) of the potential energy surface were located by minimizing the total potential energy $\mathcal{U}(\{\vec{r}_i\})$ starting from independent, instantaneous fluid configurations. Thanks to the large system size, a relatively small number of independent configurations (20--40) was sufficient to obtain good quality data. Some larger data sets (up to 300 configurations) were considered at specific state points. The LBFGS algorithm~\cite{liu__1989} was used to locate the local minima. We found that convergence becomes particularly slow close to the clustering crossover $\rho_c$ (see Sec.~\ref{sec:results}). Depending on density and temperature, from 1000 to 50000 iterations were necessary to reduce the norm of the total force per particle $W=\frac{1}{N}\sum f_i^2$ to values smaller than $10^{-13}$. Such a tolerance criterion was sufficient to reduce to zero the number of unstable modes of the dynamical matrix. 

\section{Results}\label{sec:results}

The aim of this work is to investigate the interplay between reentrant anomalies and clustering in nearly Gaussian particles, i.e. for values of $n$ slightly larger than 2.
To understand how the phase diagram is modified as the interactions deviate from the pure Gaussian potential, we start with a preliminary exploration of fluid structure of the model by varying systematically the exponent $n$. 
For this purpose, we employ the integral equation theory using the HNC closure relation, which provides a good description of the structure of soft particles at sufficiently high density~\cite{likos_effective_2001,mladek_clustering_2007}. 

A simple way to highlight the anomalous features of the structure in soft core particles is to identify the locus of state points in the $T-\rho$ plane for which the height $S^*=S(k^*)$ of the first peak of the structure factor is constant. 
We found that HNC suffers from convergence problems at intermediate densities, in particular close to the clustering crossover (see below), when $n<2.05$.
Therefore, we restrict ourselves to a relatively small value of $S^*$.
For the models at hand, we use the condition $S^*=2.3$, which is smaller than typical values observed at melting in simple liquids ($S^*\approx 2.8$ according to the Hansen-Verlet criterion~\cite{Hansen_Verlet_1969}).
We emphasize that the aim of this preliminary analysis is to highlight the presence of a structural anomaly in the $(T,\rho)$ diagram and not to track closely the fluid-solid phase boundary.
Such an operation would require adjusting the value of $S^*$ depending on density, since the first peak of $S(k)$ reaches large values (i.e., $S^*\agt 4.0$) along the cluster crystallization line~\cite{likos_why_2007,mladek_clustering_2007}.

\begin{figure}[t]
\includegraphics[width=\onefig]{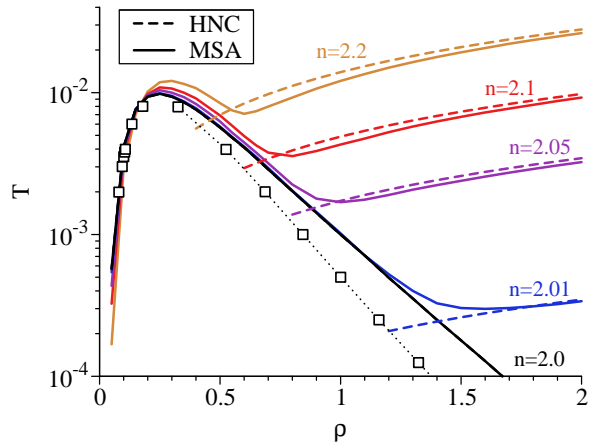}
\caption{\label{fig:iet} Integral equation theory predictions for the iso-$S^*$ lines in the $T$-$\rho$ plane for various values of $n$. 
The lines connect state points for which the height, $S^*$, of the first peak of the structure factor is equal to 2.3. 
Full and dashed lines correspond to HNC and MSA approximations, respectively. 
Open squares indicate the fluid-solid phase boundary of the GCM ($n=2$) as determined in Refs.~\cite{prestipino_phase_2005} and ~\cite{Ikeda_therm_2011}. 
For clarity, the iso-$S^*$ lines predicted by  MSA are not shown below the fluid-solid phase boundary.}
\end{figure}

The iso-$S^*$ curves obtained within the HNC approximation are shown in Fig.~\ref{fig:iet} for several values of $n \ge 2$, along with the fluid-solid phase boundary of the GCM determined in Refs.~\cite{prestipino_phase_2005} and~\cite{Ikeda_therm_2011}. 
At low density, HNC predicts the iso-$S^*$ line to depend very little on $n$ and to follow the low density branch of the GCM phase boundary.
In this portion of the phase diagram, the HNC predictions are semi-quantitative at best, since we expect the Hansen-Verlet criterion to work well along the low density phase boundary.
Thus, larger values of $S^*$ would be expected.
For all the values of $n$ shown in the figure, the iso-$S^*$ lines show a first decrease upon increasing density, following the non-monotonic, anomalous behavior of the melting line of the GCM. 
Upon increasing the density even further, an additional branch appears: for values of $n$ strictly larger than 2, the iso-$S^*$ lines follow a common envelope until they reach a crossover density and start bending upwards. 
The crossover density diverges when $n \rightarrow 2^+$, as it can be inferred by tracking the location of the minimum of $S^*(\rho)$ as a function of $n$ (not shown). 
Such a behavior is a striking confirmation of the mean-field argument of Likos et al.~\cite{Likos_Lang_Watzlawek_Lowen_2001}.
In fact, as it will be clear from the following, the minimum of the iso-$S^*$ lines marks the onset of clustering in the \textit{fluid} phase.
We find that the two coexisting anomalies are visible up to values of $n$ around 3, at which the minimum transforms into a small inflection. 
Above $n=3$ the iso-$S^*$ lines predicted by HNC display a purely monotonic increase with density.

Results obtained using the MSA are also included in the figure.
In this case, the iso-$S^*$ lines can be obtained analytically and are given by straight lines in the $T$-$\rho$ plane
$$
T =  - \frac{v^*S^*}{S^*-1} \rho 
$$
where $v^*$ is the value of the minimum of the Fourier transform of the potential. 
We see that the MSA predictions follow closely those obtained within HNC at high density.
However, the crossover between the low and intermediate density regimes (reentrant anomaly) is not captured within the mean-field description of MSA, as the latter is only reliable at high temperatures and densities.

For values of the softness exponents $n$ between $2^+$ and approximately 3, HNC thus predicts the existence of three regimes in the fluid phase diagram separated by two anomalies:
(i) the ``reentrant'' anomaly well known from studies of the GCM and Hertzian spheres, corresponding to a maximum in the iso-$S^*$ lines, 
and (ii) a ``cluster'' anomaly, observed as the particles start to explore the ultrasoft portion of the potential, which appears as a minimum in the iso-$S^*$ lines.
To test this anomalous scenario in-depth, we focus in the following on the case $n=2.2$, for which we collected the most extensive set of simulation data.
Simulations for different values of $n$ will be briefly discussed at the end of this section.

\begin{figure}[t]
\includegraphics[width=\onefig]{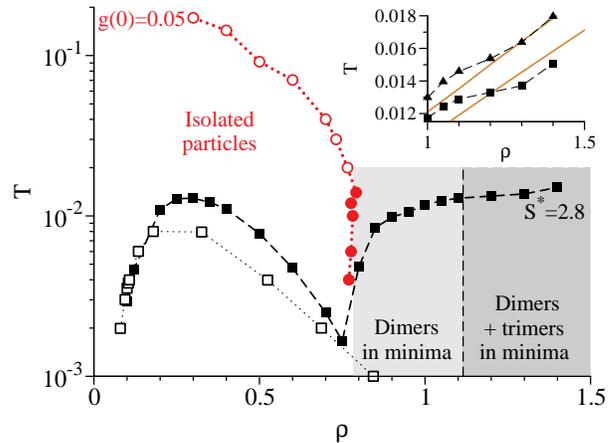}
\caption{\label{fig:phase_diagram} Fluid phase diagram of the GEM-2.2 in the $T$-$\rho$ plane. 
State points at which $S^*=2.8$ are indicated as full squared and are connected by full lines (iso-$S^*$ lines).
State points at which $g(0)=0.05$ are drawn as filled or open circles according to whether they are characterized by clustering or transient overlaps, respectively, as explained in the text.
Open squares indicate the fluid-solid phase boundary of the GCM ($n=2$) as determined in Refs.~\cite{prestipino_phase_2005} and ~\cite{Ikeda_therm_2011}.
The light grey and dark grey shaded areas indicate portions of the phase diagram displaying finite fractions of dimers and trimers, respectively, in the local minima of the potential energy surface.
Inset: magnification of the behavior of the iso-$S^*$ line at large densities for $S^*=2.8$ (squares) and $S^*=2.3$ (triangles).
Also included are the corresponding results from HNC (solid lines).
}
\end{figure}

\begin{figure}[t]
\includegraphics[width=\onefig]{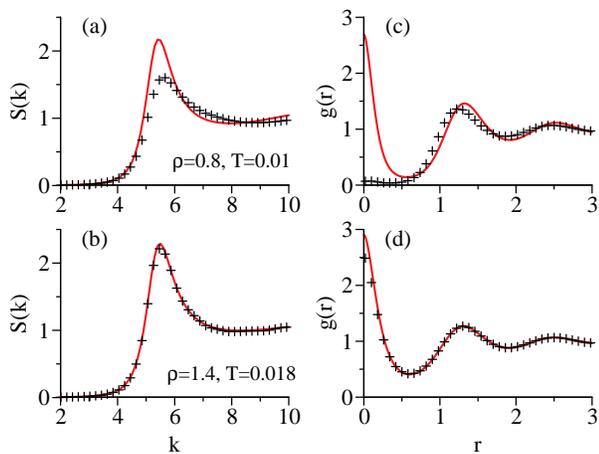}
\caption{\label{fig:iet_md} Comparison of the structure factors $S(k)$ [panels (a) and (b)] and  radial distribution functions $g(r)$ [panel (c) and (d)] obtained using the HNC closure (solid lines) and MD simulations (crosses). Data are shown for $\rho=0.8, T=0.01$ in (a) and (c), and for $\rho=1.4, T=0.018$ in (b) and (d).} 
\end{figure}

In Fig.~\ref{fig:phase_diagram} we combine our molecular dynamics results and the analysis of the potential energy surface to characterize the fluid phase diagram the GEM-2.2.
We start by discussing the shape of the iso-$S^*$ line, which we determined using the value $S^*=2.8$ (Hansen-Verlet criterion).
At low density, the line first follows the trend of the GCM fluid-solid phase boundary.
Upon further compression, it first passes through a maximum located around $\rho_r\approx 0.3$ and then a minimum around  $\rho_c\approx 0.78$.
Thus both the reentrant and cluster anomalies predicted by HNC are visible in the simulation data.
Upon closer inspection, however, it is clear that the agreement between HNC and the simulations in this portion of the phase diagram is only qualitative.
The minimum separating the soft-core and the cluster regimes is in fact deeper than predicted by HNC.
These discrepancies are further highlighted in Figs.~\ref{fig:iet_md}(a) and (c), where we compare the HNC predictions for the structure factor $S(k)$ and the radial distribution function $g(r)$ around the crossover.
HNC predicts clustering to occur at smaller densities (or higher temperatures) than actually observed in the simulations, as seen from the differences in $g(r)$ around $r=0$.
Only at larger densities and temperatures [see Figs.~\ref{fig:iet_md}(b) and (d)], the HNC predictions agree quantitatively with the simulations, as expected from previous work~\cite{mladek_clustering_2007}.
This is further illustrated by the iso-$S^*$ lines shown in the inset of Fig.~\ref{fig:phase_diagram}, where we observe good agreement between HNC and simulations for $S^*=2.3$, but not for $S^*=2.8$.
We conclude that better closure relations are required for a quantitative description of the fluid phase diagram close to the fluid-solid phase boundary, especially at low and intermediate densities.

\begin{figure*}[t]
\includegraphics[width=\twofig]{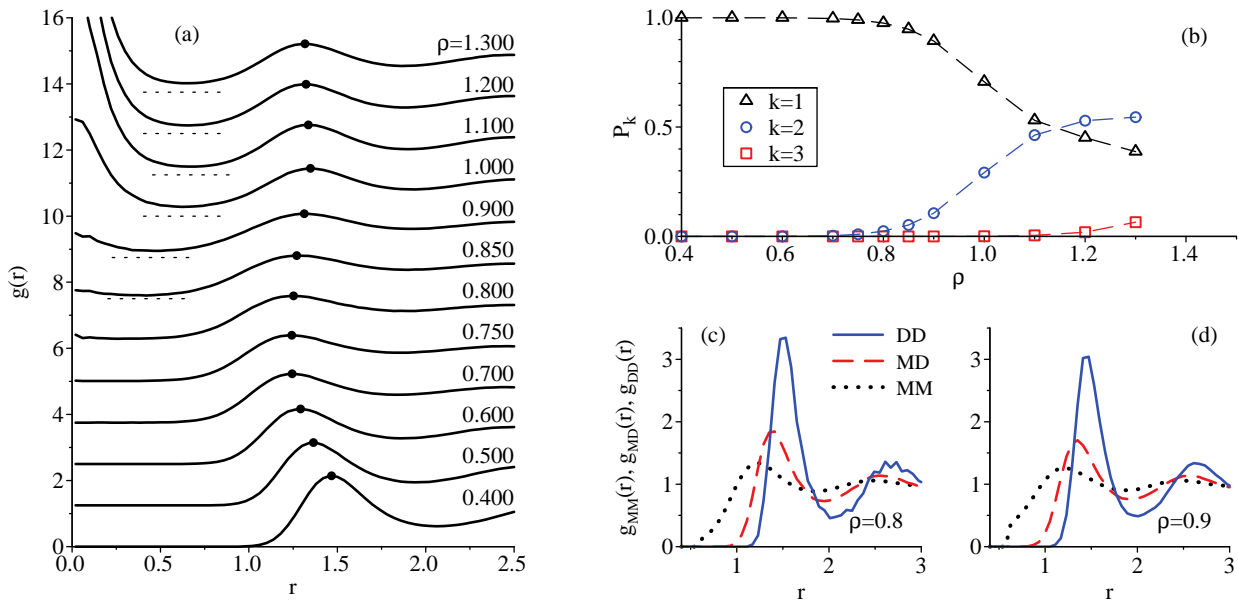}
\caption{\label{fig:inst} Density variation of the radial distribution function and cluster properties of the GEM-2.2 evaluated for instantaneous fluid configurations:
(a) $g(r)$ along the isotherm $T=0.012$.
For clarity, each line has been vertically shifted from the previous one  by a constant offset, 1.25.
For $\rho>0.8$, the horizontal dotted lines indicate the ordinates at which $g(r)$ would equal zero.
Filled circles indicate the first neighbor shell peaks.
(b) Fraction of $k$-mers for $T=0.012$  as a function of $\rho$ for $k=1$ (monomers, triangles), 2 (dimers, circles), and 3 (trimers, squares).
(c) and (d) Monomer-monomer (dotted  line), monomer-dimer (dashed line) and dimer-dimer (solid line) radial distribution functions evaluated at (c) $\rho=0.8$, $T=0.012$ and (d) $\rho=0.9$, $T=0.012$. 
Note that the distribution functions are zero by construction for $r<d=0.6$.
}
\end{figure*}

To illustrate the connection between the minimum of the iso-$S^*$ line and clustering, we evaluate the radial distribution function $g(r)$ and analyze its isothermal variation. 
In the following, we will denote with $r^*$ the position of the peak of $g(r)$ corresponding to the first neighbor (FN) shell.
In Fig.~\ref{fig:inst}(a) we show the $g(r)$ along the isotherm $T=0.012$.
As the density is increased from $\rho=0.4$, the first peak initially shifts markedly towards smaller distances, as expected for soft-core particles.
Upon crossing $\rho_c$, a broad peak develops at $r=0$ and grows by further increasing the density.
The presence of a well-defined minimum between $r=0$ and $r^*$, indicates the formation relatively stable clusters.
The FN peak position $r^*$ also shows an interesting behavior: it first decreases with increasing density up to $\rho_c$, then it starts to increase until the density reaches approximately $\rho=1.1$.
This point will be further discussed below.

In contrast to cluster formation, transient overlaps can always occur at any density in fluids of ultrasoft particles as a consequence of thermal fluctuations.
Indeed, at high temperatures, coreless particles often superpose during collisions, which implies a finite value of $g(r=0)$.
How to distinguish between clustering and such transient overlaps?
We propose to classify clustering states as those for which the second derivative $g^{\prime\prime}(r=0)<0$.
In practice, an equivalent and numerically more stable criterion is to check whether the $g(r)$ displays a minimum between 0 and $r^*$.
States for which $g(0)$ is finite but $g^{\prime\prime}(0)>0$ are characterized instead by transient overlaps.

With all this information at hand we return to the fluid phase diagram Fig.~\ref{fig:phase_diagram} and mark the boundary between normal and cluster states.
To this end, we identify the state points in the $T$-$\rho$ plane at which $g(0)$ reaches constant, ``small'' value.
The reference value $g(0)=0.05$ is small enough to detect the early onset of clusters and overlaps, still sufficiently large to minimize statistical uncertainties.
Numerically, we evaluate $g(0)$ from the probability of finding particles at distances less than 0.03.
Following the classification introduced above,  we use open or filled circles in Fig.~\ref{fig:phase_diagram} for   states characterized by clusters or transient overlaps, respectively.
In the low and intermediate density regime, the onset of transient overlaps is located at temperatures well above the melting line.
Upon increasing the density, the iso-$g(0)$ line bends downwards and starts to run vertically in the $T-\rho$ plane.
In this portion of the phase diagram, the sign of $g^{\prime\prime}(0)$ along iso-$g(0)$ line becomes negative, signaling the transition to cluster states.
By visual inspection, it is clear that the boundary between normal and cluster states corresponds well to the minimum of the iso-$S^*$ line, which motivates the name ``cluster anomaly''.

To understand the physical origin of the cluster anomaly, we analyze in more detail the changes in the microscopic structure of the clusters.
To identify clusters, we use the same geometrical definition as in previous work~\cite{coslovich_hopping_2011,coslovich_cluster_2012}: a particle is considered to be part of a cluster if its distance to at least another particle of that cluster is smaller than a fixed value $d$. 
Typically, $d$ is chosen around the first minimum of $g(r)$, which is located in the range 0.6--0.7 over the pertinent range of density and temperatures.
For simplicity, in our analysis we used a state independent cut off $d=0.6$.

The cluster population $P_k=\langle N_k\rangle / N$, where $N_k$ is the number of $k$-mers in a configuration, is shown in  Fig.~\ref{fig:inst}(b) as a function of density along the isotherm $T=0.012$.
We find that starting from $\rho\approx 0.78$ the fluid is no longer composed only of isolated particles (monomers) but possesses a growing fraction of dimers.
In the range of density between  0.78 and 1.1, we observe coexistence between monomers and dimers, the fraction of  higher order $k$-mers being zero within our numerical accuracy.
We remark that this range varies slightly as a function of the chosen temperature.
To illustrate the real space structure of coexisting monomers and dimers, we show in Fig~\ref{fig:inst}(c) and (d) the monomer-monomer (MM), monomer-dimer (MD) and dimer-dimer (DD) radial distribution functions evaluated at $\rho=0.8$ and $\rho=0.9$, respectively.
These correlation functions are calculated from
$$
g_{\alpha\beta}(r) = \frac{V}{\langle N_\alpha\rangle \langle N_\beta\rangle} \left\langle \sum_{i=1}^{N_\alpha}\sum_{j=1}^{N_\beta}\delta(\vec{r} - \vec{r}_{ij})\right\rangle
$$
where $\alpha$ and $\beta$ are either $M$ or $D$ and the term $i=j$ in the double sum is ignored when $\alpha=\beta$.
For dimers, the distance $\vec{r}_{ij}$ is evaluated from their center of mass.
We find that monomers are more loosely correlated than dimers, as it can be evinced from the nearly flat shape of $g_\textrm{MM}(r)$.
This probably reflects the fact the effective interactions between dimers are harsher than the bare potential between individual particles.
Dimers, on the other hand, tend to occupy a larger volume and are thus separated from their neighbors by larger distances.
This suggests that the slight increase of $r^*$ above $\rho_c$ is due to the larger FN shell of the growing dimers population.

Upon further compression above $\rho=1.1$, a small fraction of trimers $P_3$ appears and grows by increasing density at the expenses of $P_1$ and $P_2$ [see Fig.~\ref{fig:inst}(b)].
Keeping $T=0.012$ fixed, we cannot extend the analysis of the fluid regime beyond $\rho\sim 1.3$, due to incipient, partial cluster crystallization.
We thus conclude that, above $\rho_c$, clusters are characterized by low population numbers, i.e. 2 or 3, and coexist in a nearly structureless continuum formed by isolated particles.
This scenario is most likely connected to the underlying crystalline phase  diagram, which may possess a cascade of transitions involving normal fluids (i.e., isolated particles) and cluster crystals with integer population numbers.
Such a cascade of complex crystal structures is known to occur in the PSM~\cite{Likos_PSM_1998,Zhang_PSM_2012} and in the GEM-4~\cite{Zhang_Charbonneau_Mladek_2010,Neuhaus_Likos_2011}.

\begin{figure*}[t]
\includegraphics[width=\twofig]{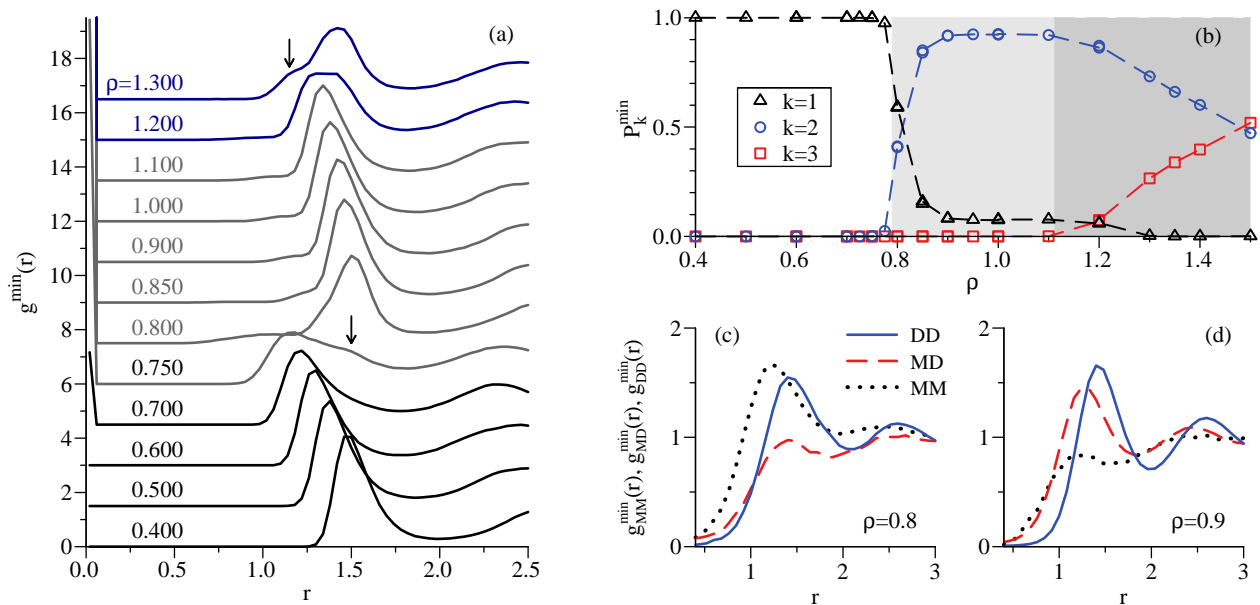}
\caption{\label{fig:min} Density variation of the radial distribution function and cluster properties of the GEM-2.2 evaluated for local minima of the potential energy surface. Lines and symbols as in Fig.~\ref{fig:inst}.
Note that in this case both the radial distribution function $g^\textrm{min}(r)$ and the cluster populations are independent of $T$ over the studied range of state parameters.
The actual data shown in the figure were obtained from minimizations at $T=0.02$.
In (a), solid lines are colored according to the different regimes of cluster populations, also indicated by the shaded areas in (b).
Vertical arrows highlight the splitting of the FN maxima at $\rho=0.75$ and $\rho=1.3$ and indicate the location of the secondary peaks.
In (c) and (d) the partial radial distribution functions are shown for $\rho=0.8$, $T=0.02$ and $\rho=0.9$, $T=0.02$, respectively.}
\end{figure*}

Due to the thermal motion of particles, however, the definition of a cluster in the fluid states is subject to some ambiguity.
For instance, the drop of $g_\textrm{MM}(r)$ at $r=d$ for $\rho=0.9$ (and to a less extent for $\rho=0.8$) indicates that there is no clear separation between the monomers and the FN shells of the clusters.
To obtain a sharper picture, we apply our cluster analysis to the local minima of the underlying potential energy surface (PES).
We perform a statistical sampling of the PES properties by quenching several independent samples (instantaneous configurations) at a given $\rho$ and $T$,
as described in the Sec.~\ref{sec:methods}. 
In the following, we will focus on the isotherm $T=0.02$ and study how the underlying landscape changes upon varying the density.
Although the PES is fully determined once the number of particles and the volume are specified, the properties  of the basins \textit{visited} by the system may vary below some onset temperature~\cite{sastry__1998}.
However, this is not observed for the GEM-2.2, at least down to temperatures close to the fluid-solid phase boundary.
The situation resembles the scenario observed in simple liquids, where the properties of local minima visited above melting are indeed temperature independent~\cite{sastry__1998}.

The main advantage over the analysis based on instantaneous configurations is that,  in the local minima, clusters  appear as much tighter conglomerates of particles, sitting practically on top of each other.
This is demonstrated by the radial distribution function $g^\text{min}(r)$ evaluated using the minimized configurations, see Fig.~\ref{fig:min}(a).
The peak at zero distance, which emerges at the clustering crossover, is now extremely narrow and is separated from the FN shell by a clear gap where $g(r)$ is zero within numerical precision.
This, in turn, makes the definition of clusters straightforward and unambiguous.
In practice, we used $d=0.2$ as a cut off distance for the cluster analysis on local minima.

Compared to instantaneous configurations, local minima present a more subtle density dependence of the radial distribution function.
The latter is best understood when discussed along with that of the cluster population $P^\textrm{min}_k$, which we show in Fig.~\ref{fig:min}(b).
An analysis of the cluster populations reveals a steep rise of dimers at $\rho_c=0.78$, making the definition of the clustering crossover very sharp.
The fraction of trimers becomes non zero only above $\rho=1.1$, again displaying a sharper growth than in instantaneous configurations. 
Upon increasing the density even further ($\rho>1.3$), monomers eventually disappear and we observe coexistence between dimers and trimers.
These various regimes are indicated in Fig.~\ref{fig:phase_diagram} by shaded areas, to highlight their connection to the observed anomalies.

We can now more easily understand the evolution of $g^\textrm{min}(r)$ with density.
The FN peak of $g^\textrm{min}(r)$ shows a broadening around $\rho_c$ and then splits upon further increasing the density.
The two peaks arise from the coexistence of monomers and dimers, which are characterized by two distinct typical inter-cluster separations.
This is confirmed by an analysis of the monomer-monomer and dimer-dimer contributions to $g^\textrm{min}(r)$, which are calculated as for instantaneous configurations and are shown in Fig.~\ref{fig:min}(c) and (d) at $\rho=0.8$ and $\rho=0.9$, respectively.
In particular, the splitting of the FN shell for $\rho=0.75$ [see panel (a)] is due to growing dimer-dimer correlations.
Indeed, the position of the secondary maximum at $r\approx 1.45$ matches the location of the first peak in $g_\textrm{DD}(r)$.
We also find that the MM peak at small distances soon looses its structure above $\rho_c$, whereas the DD peak shifts progressively to smaller distances.
This, together with the density evolution of $g^\textrm{min}(r)$ above $\rho_c$, suggests that dimers behave as larger, effective soft-core particles.
Above $\rho=1.1$, the FN peak of $g^\textrm{min}(r)$ shows again a broadening and a subsequent splitting, which arises now from the coexistence of subpopulations of dimers and trimers in the local minima.

The partial radial distribution functions $g^\textrm{min}_{\alpha\beta}(r)$ for $\rho=0.9$ [see Fig.~\ref{fig:min}(d)]  qualitatively confirm the results obtained from instantaneous configurations in the monomer-dimer coexistence region.
In particular, we see that isolated particles, which amount to $\approx 10\%$ of the $k$-mers at this density, are spatially uncorrelated, as indicated by the unstructured profile of $g_\textrm{MM}(r)$.
Very close to the cluster crossover [see Fig.~\ref{fig:min}(c)], we observe a strong suppression of $g^\textrm{min}_\textrm{MD}(r)$ and and a concomitant enhancement of monomer-monomer correlations.
This indicates that monomers and dimers have a tendency to microsegregate and suggests an underlying instability with respect to phase separation into coexisting fluid and cluster phases.
Visual inspection of the real space structure of the local minima at $\rho=0.8$ confirms these observations.
We found no such tendency to microsegregation in the instantaneous configurations.
These results resemble the scenario observed in certain phase separating binary Lennard-Jones mixtures~\cite{sarkar_inherent_2011}, for which the local minima display clear signs of demixing---even for states at which the equilibrium fluid is essentially homogeneous.

\begin{figure}[t]
\includegraphics[width=\onefig]{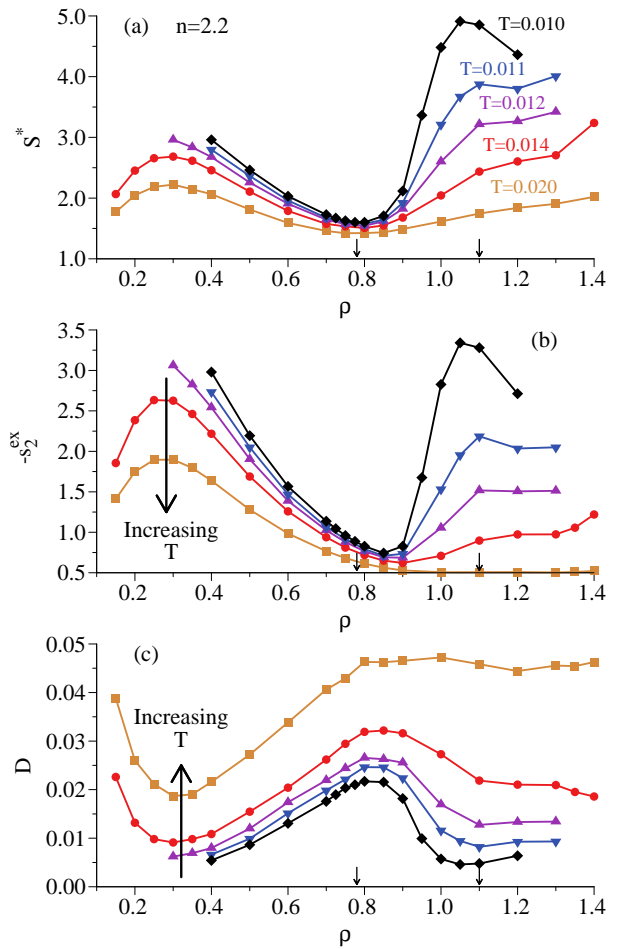}
\caption{\label{fig:anomaly2.2} Structural and dynamic anomalies of the GEM-2.2 as a function of density along selected isotherms: $T=0.02$ (squares), 0.014 (circles), 0.012 (triangles), 0.011 (reversed triangles), and 0.01 (diamonds).
(a) Height, $S^*$, of the first peak of the structure factor. (b) Inverse of two-body excess entropy $s^\textrm{ex}_2$.
(c) Diffusion constant $D$. Arrows at $\rho=0.78$ and $\rho=1.1$ mark the appearance of dimers and trimers in the PES, respectively.
}
\end{figure}

We now analyze the connection between the structural anomalies and the dynamics.
It is well established~\cite{krekelberg_anomalous_2009} that the reentrant anomaly in soft-core models is accompanied by a dynamic anomaly in the single particle diffusivity: 
beyond the reentrant crossover, the dynamics accelerates upon increasing density due to the progressive the loss of structural correlations. 
We  now show that also the cluster  anomalies also have their dynamic counterparts.
In the three panels of Fig.~\ref{fig:anomaly2.2} we compare the isothermal variation of the peak height $S^*$ of the structure factor, the inverse of the two-body contribution to the excess entropy
$$
s^\textrm{ex}_2 = - 2\pi \rho \int_0^\infty r^2\left\{ g(r) \ln g(r) - [g(r) -1]\right\} dr
$$
and the diffusion constant $D$, which has been obtained from the usual Einstein relation
$$
\langle R^2(t) \rangle \rightarrow 6Dt
$$
where $R^2(t) = \frac{1}{N} \sum_{i=1}^{N}|\vec{r}_i(t) - \vec{r}_i(0)|^2$ is the mean square displacement. 
The two-body excess entropy $s^\textrm{ex}_2$ is a simple measure of pair translation order in the system.
This quantity has been used to characterize structural anomalies in fluids of ultrasoft particles~\cite{krekelberg_anomalous_2009,krekelberg_generalized_2009}.

\begin{figure}[t]
\includegraphics[width=\onefig]{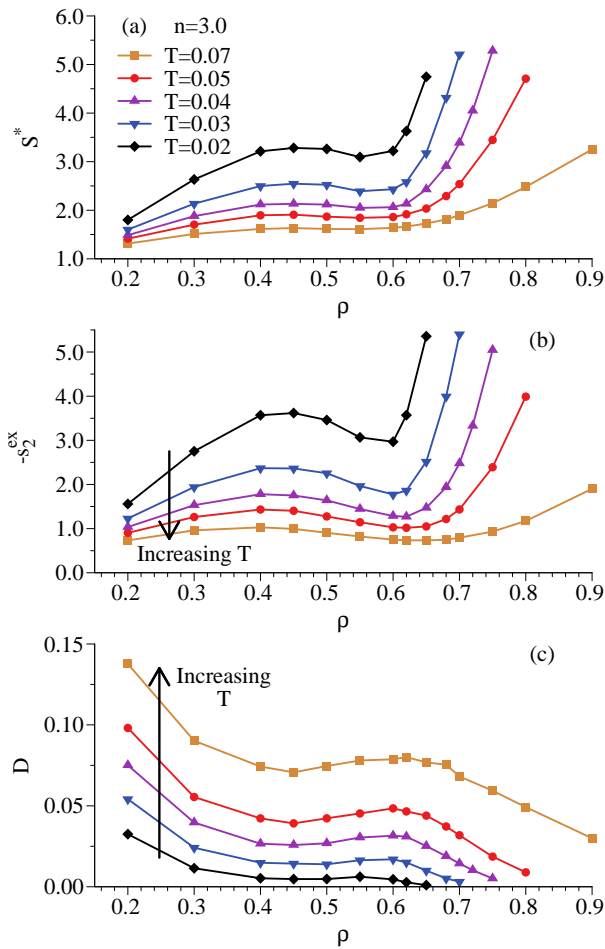}
\caption{\label{fig:anomaly3.0}Structural and dynamic anomalies of the GEM-3 as a function of density along selected isotherms: $T=0.07$ (squares), 0.05 (circles), 0.04 (triangles), 0.03 (reversed triangles), and 0.02 (diamonds).
(a)  $S^*$,  of the first peak of the structure factor. (b) Inverse of the two-body excess entropy $s^\textrm{ex}_2$.
(c) Diffusion constant $D$.
}
\end{figure}

The two isotherms at high temperature reproduce the reentrant anomalies in both structure and dynamics, as expected.
On the contrary the two low temperature isotherms necessarily hit, at small $\rho$, on the fluid-bcc phase boundary, and therefore only the cluster anomalies are visible.
We find that $s^\textrm{ex}_2$ follows closely the density variation of the $S^*$.
A close inspection of panels (a) and (b) of Fig.~\ref{fig:anomaly2.2} reveals only a minor discrepancy between the location of the minima in $S^*$ and $s^\textrm{ex}_2$, with the latter providing a slightly larger ($\approx 5\%$) value of the density at the minimum.
A comparison of the structural metrics and $D$ shows that the clustering crossover at $\rho_c$ is clearly accompanied by a decrease in diffusivity.
Physically, this anomaly can be easily understood, since transient clusters temporarily behave as effective, heavier particles, thus slowing down the dynamics of the fluid.
Despite the clear connection between structure and dynamics illustrated by Fig.~\ref{fig:anomaly2.2}, an attempt to collapse the dynamic data as a function of the two-body excess entropy $s^\textrm{ex}_2$, as in previous works~\cite{krekelberg_anomalous_2009,krekelberg_generalized_2009}, did not yield satisfactory results on a quantitative level (see also~\cite{Fomin_Ryzhov_Gribova_2010} for a discussion on the breakdown of such a ``Rosenfeld scaling''~\cite{rosenfeld_quasi-universal_1999}).

\begin{figure}[t]
\includegraphics[width=\onefig]{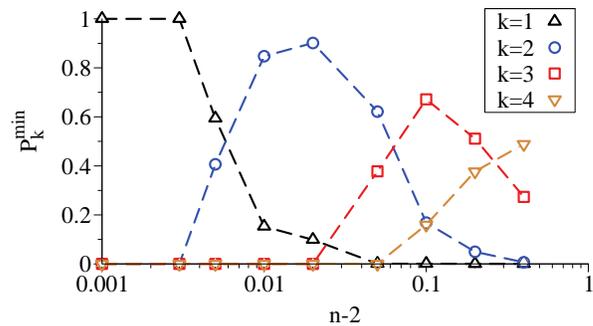}
\caption{\label{fig:Pncl_alpha} Fraction of $k$-mers in local minima of the potential energy surface as a function of $n-2$ for $k=1$ (triangles), 2 (circles), 3 (squares), and 4 (reversed triangles).
Local minima were sampled at $\rho=2.0$, $T=0.1$ .
}
\end{figure}

Around $\rho \approx 1.1$ and at sufficiently low temperatures, an additional anomaly is visible as a maximum in $S^*$ (and $s^\textrm{ex}_2$) and a minimum in $D$.
The relatively large values attained by $S^*$ along the lowest isotherm, $T=0.01$, suggest that the system might be slightly supercooled, despite being (meta)stable during the whole length of our simulations.
Figure~\ref{fig:anomaly2.2}(a) and (b) show however that this maximum  develops smoothly by decreasing temperature.
The inflection of the iso-$S^*$ lines in the $(T,\rho)$ diagram [see inset of Fig.~\ref{fig:phase_diagram}] is most likely a high temperature manifestation of this additional anomaly.
Thus, we are confident that this represents an actual feature of the \textit{equilibrium} fluid.
This anomalous scenario is also observed at smaller $n$ values, down to at least 2.01 (not shown).
In this case, the two cluster anomalies are visible at much lower temperatures than in the GEM-2.2, as expected from the HNC results.

Qualitatively, the second cluster anomaly in the GEM-2.2 correlates with the appearance of trimers and the disappearance of isolated particles.
However, its underlying physical origin remains unclear:
for instance, why should the formation of trimers accelerate the dynamics?
A possible explanation is that, at these densities, different subpopulations of $k$-mers coexist ($k=1,2,3$), which enhances the structural disorder and thus accelerates the dynamics.
It could also be argued that dimers behave as effective, ultrasoft particles and display an additional reentrant anomaly in their own right.
We note, in passing, that neither HNC nor MSA can reproduce this anomalous feature.
Further work is definitely needed to clarify this issue.

By varying the softness exponent, we found that coexistence of reentrant and cluster anomalies is well visible up to at least $n=3$, a value that can be used to model the effective interactions between amphiphilic dendrimers~\cite{mladek_computer_2008}.
This is demonstrated in Fig.~\ref{fig:anomaly3.0}, where we show the isothermal variation of $S^*$, $s^\textrm{ex}_2$ and $D$ in the GEM-3.
In this case, however, only the first cluster anomaly is visible in the range of densities and temperatures covered by our simulations.
The absence of the second anomaly is consistent with the fact the fraction of trimers in the fluid is negligible ($P_3<10^4$) at the state points shown in Fig.~\ref{fig:anomaly3.0}.

Finally, we study the evolution of the cluster population in the local minima as $n\rightarrow 2^+$.
Figure~\ref{fig:Pncl_alpha} shows the fraction of $k$-mers $P_k$ with $k=1,2,3,4$ at fixed $\rho=2.0$ and $T=0.1$.
As the interaction smoothly transforms into a Gaussian, clusters of high order disappear progressively.
At the studied density, only models with $n-2 \leq 0.003$ are fully dissociated within our statistics.
The variation of $P_k$ with $n$ thus provides a striking confirmation of the Likos criterion for clustering in fluid states.

\section{Conclusions}\label{sec:conclusions}

In this work we demonstrated the existence of multiple anomalies in the structure and dynamics of fluids composed of nearly Gaussian particles.
We focused in particular on particles interacting via the generalized exponential potential, Eq.~\eqref{eqn:u}, with an exponent  $n$ slightly larger than 2.
Our most extensive set of data concerns the GEM-2.2, for which we gathered molecular dynamics results and for which we analyzed systematically the potential energy surface.

At low densities, the structure and dynamics of the fluid display a non-monotonic density variation. 
Such a reentrant anomaly is well-known in soft-core particles~\cite{krekelberg_anomalous_2009,Pamies_Cacciuto_Frenkel_2009} and manifests itself as a loss of translation order upon compression and a concomitant acceleration of the single-particle dynamics.
Nearly Gaussian particles show two additional anomalies by compressing the fluid at constant temperature:
the height $S^*$ of the first peak of the structure factor displays a minimum and an additional inflection at higher densities.
At sufficiently low temperatures, such an inflection transforms into a small but well-defined minimum.
All these anomalies have counterparts in the single particle dynamics, with minima of $S^*$ corresponding to maxima of the diffusion constant $D$, and viceversa.
We note that a simple scaling between Newtonian and overdamped Langevin dynamics has been recently reported for ultrasoft fluids~\cite{Pond_Errington_Truskett_2011}, indicating that the connection between structural and dynamical anomalies observed here is likely quite general.

At  high densities, particles explore the bounded part of the potential.
The two additional anomalies occurring in this portion of the fluid phase diagram, can thus be attributed to the appearance of clusters in the fluid.
Namely, we found that the minimum in $S^*(\rho)$  correlates with   the sharp growth of dimers in both instantaneous configuration and in the local minima of the PES.
The additional maximum ostensibly correlates to the appearance of trimers  upon further compression, although its actual physical origin needs to be clarified.
Interestingly, the spatial structure of the local minima reveals that coexisting subpopulations of isolated particles and clusters can show a slight segregation.
This suggests an underlying instability of fluid with respect to phase separation into coexisting fluid and cluster phases.
It would thus be interesting to characterize in more detail the low temperature phase diagram of nearly Gaussian particles, along the lines of recent numerical work on the GEM-4~\cite{Zhang_Charbonneau_Mladek_2010}.

We demonstrated that clustering in the fluid phase disappears in a continuous way as the interaction potential tends to a purely Gaussian shape, i.e., as the softness exponent $n$ approaches 2 from above.
This is an explicit confirmation of the mean-field argument by Likos et al.~\cite{Likos_Lang_Watzlawek_Lowen_2001}, which has never been so far tested explicitly on GEMs with $n$ close to 2.
What our results clarify, however, is that reentrant and clustering behavior are not mutually exclusive, as is sometimes implicit in the presentation of $Q^+$ and $Q^{\pm}$ models~\cite{Likos_Lang_Watzlawek_Lowen_2001}.
We found that coexistence of reentrant and cluster anomalies is well visible up to at least $n=3$.
This behavior was anticipated by Zhang et al~\cite{Zhang_Charbonneau_Mladek_2010} in their study of the low temperature phase diagram of the GEM-4 model and has been confirmed here more systematically.
Whether these coexisting anomalies may be observed in colloidal suspensions of highly branched macromolecules, such as dendrimers, is an open question that awaits an experimental verification.

Another interesting implication of our work concerns the nature of the possible glassy states of ultrasoft particles.
It has been shown that the GCM becomes a surprisingly good glass-former at sufficiently high densities ($\rho \sim 2.0$)~\cite{ikeda_glass_2011}.
This has been attributed to a frustration effect induced by the long range of the Gaussian potential. 
We found that for $n=2.01$ the cluster anomaly is shifted up to approximately $\rho\approx 2.0$.
This, in turn, opens up a broad region in the fluid phase diagram in which the model behaves essentially as the GCM.
In such a region, models with $n$ values marginally larger than 2 might display peculiar glass-forming properties.
In particular, it would be interesting to see whether the proximity to cluster formation destabilizes the glassy behavior of the GCM and to what extent such a behavior will be affected by clustering at even higher densities.
Finally, we believe that the analysis of the PES may prove particularly useful in the study of amorphous cluster phases of size dispersed ultrasoft particles~\cite{coslovich_cluster_2012}.
Work to investigate the properties of the PES explored by cluster glass-forming fluids is currently underway.

\begin{acknowledgements} 

We acknowledge the HPC@LR Center of Competence in High-Performance Computing of Languedoc-Roussillon (France) for allocation of CPU time.
A. I. acknowledges the financial support from JSPS postdoctral fellowship for research abroad.

\end{acknowledgements}


\begin{thebibliography}{10}%
\makeatletter
\providecommand \@ifxundefined [1]{%
 \ifx #1\undefined \expandafter \@firstoftwo
 \else \expandafter \@secondoftwo
\fi
}%
\providecommand \@ifnum [1]{%
 \ifnum #1\expandafter \@firstoftwo
 \else \expandafter \@secondoftwo
\fi
}%
\providecommand \enquote [1]{``#1''}%
\providecommand \bibnamefont  [1]{#1}%
\providecommand \bibfnamefont [1]{#1}%
\providecommand \citenamefont [1]{#1}%
\providecommand\href[0]{\@sanitize\@href}%
\providecommand\@href[1]{\endgroup\@@startlink{#1}\endgroup\@@href}%
\providecommand\@@href[1]{#1\@@endlink}%
\providecommand \@sanitize [0]{\begingroup\catcode`\&12\catcode`\#12\relax}%
\@ifxundefined \pdfoutput {\@firstoftwo}{%
 \@ifnum{\z@=\pdfoutput}{\@firstoftwo}{\@secondoftwo}%
}{%
 \providecommand\@@startlink[1]{\leavevmode}%
 \providecommand\@@endlink[0]{}%
}{%
 \providecommand\@@startlink[1]{%
  \leavevmode
  \pdfstartlink
   attr{/Border[0 0 1 ]/H/I/C[0 1 1]}%
   user{/Subtype/Link/A<</Type/Action/S/URI/URI(#1)>>}%
  \relax
 }%
 \providecommand\@@endlink[0]{\pdfendlink}%
}%
\providecommand \url  [0]{\begingroup\@sanitize \@url }%
\providecommand \@url [1]{\endgroup\@href {#1}{\urlprefix}}%
\providecommand \urlprefix [0]{URL }%
\providecommand \Eprint[0]{\href }%
\@ifxundefined \urlstyle {%
  \providecommand \doi [1]{doi:\discretionary{}{}{}#1}%
}{%
  \providecommand \doi [0]{doi:\discretionary{}{}{}\begingroup
  \urlstyle{rm}\Url }%
}%
\providecommand \doibase [0]{http://dx.doi.org/}%
\providecommand \Doi[1]{\href{\doibase#1}}%
\providecommand \bibAnnote [3]{%
  \BibitemShut{#1}%
  \begin{quotation}\noindent
    \textsc{Key:}\ #2\\\textsc{Annotation:}\ #3%
  \end{quotation}%
}%
\providecommand \bibAnnoteFile [2]{%
  \IfFileExists{#2}{\bibAnnote {#1} {#2} {\input{#2}}}{}%
}%
\providecommand \typeout [0]{\immediate \write \m@ne }%
\providecommand \selectlanguage [0]{\@gobble}%
\providecommand \bibinfo [0]{\@secondoftwo}%
\providecommand \bibfield [0]{\@secondoftwo}%
\providecommand \translation [1]{[#1]}%
\providecommand \BibitemOpen[0]{}%
\providecommand \bibitemStop [0]{}%
\providecommand \bibitemNoStop [0]{.\EOS\space}%
\providecommand \EOS [0]{\spacefactor3000\relax}%
\providecommand \BibitemShut [1]{\csname bibitem#1\endcsname}%
\bibitem{Likos_2006}%
  \BibitemOpen
  \bibfield{author}{%
  \bibinfo {author} {\bibfnamefont{C.~N.}\ \bibnamefont{Likos}},\ }%
  \bibfield{journal}{%
  \bibinfo {journal} {Soft Matter}\ }%
  \textbf{\bibinfo {volume} {2}},\ \bibinfo {pages} {478} (\bibinfo {year}
  {2006})%
  \bibAnnoteFile{NoStop}{Likos_2006}%
\bibitem{Vlassopoulos_Cloitre_2012}%
  \BibitemOpen
  \bibfield{author}{%
  \bibinfo {author} {\bibfnamefont{D.}~\bibnamefont{Vlassopoulos}}\ and\
  \bibinfo {author} {\bibfnamefont{M.}~\bibnamefont{Cloitre}},\ }%
  \bibfield{journal}{%
  \bibinfo {journal} {Soft Matter}\ }%
  \textbf{\bibinfo {volume} {8}},\ \bibinfo {pages} {4010} (\bibinfo {year}
  {2012})%
  \bibAnnoteFile{NoStop}{Vlassopoulos_Cloitre_2012}%
\bibitem{Likos1998}%
  \BibitemOpen
  \bibfield{author}{%
  \bibinfo {author} {\bibfnamefont{C.~N.}\ \bibnamefont{Likos}}, \bibinfo
  {author} {\bibfnamefont{H.}~\bibnamefont{L\"owen}}, \bibinfo {author}
  {\bibfnamefont{M.}~\bibnamefont{Watzlawek}}, \bibinfo {author}
  {\bibfnamefont{B.}~\bibnamefont{Abbas}}, \bibinfo {author}
  {\bibfnamefont{O.}~\bibnamefont{Jucknischke}}, \bibinfo {author}
  {\bibfnamefont{J.}~\bibnamefont{Allgaier}},\ and\ \bibinfo {author}
  {\bibfnamefont{D.}~\bibnamefont{Richter}},\ }%
  \bibfield{journal}{%
  \bibinfo {journal} {Phys. Rev. Lett.}\ }%
  \textbf{\bibinfo {volume} {80}},\ \bibinfo {pages} {4450} (\bibinfo {year}
  {1998})%
  \bibAnnoteFile{NoStop}{Likos1998}%
\bibitem{zaccarelli_tailoring_2005}%
  \BibitemOpen
  \bibfield{author}{%
  \bibinfo {author} {\bibfnamefont{E.}~\bibnamefont{Zaccarelli}}, \bibinfo
  {author} {\bibfnamefont{C.}~\bibnamefont{Mayer}}, \bibinfo {author}
  {\bibfnamefont{A.}~\bibnamefont{Asteriadi}}, \bibinfo {author}
  {\bibfnamefont{C.~N.}\ \bibnamefont{Likos}}, \bibinfo {author}
  {\bibfnamefont{F.}~\bibnamefont{Sciortino}}, \bibinfo {author}
  {\bibfnamefont{J.}~\bibnamefont{Roovers}}, \bibinfo {author}
  {\bibfnamefont{H.}~\bibnamefont{Iatrou}}, \bibinfo {author}
  {\bibfnamefont{N.}~\bibnamefont{Hadjichristidis}}, \bibinfo {author}
  {\bibfnamefont{P.}~\bibnamefont{Tartaglia}}, \bibinfo {author}
  {\bibfnamefont{H.}~\bibnamefont{L�wen}},\ and\ \bibinfo {author}
  {\bibfnamefont{D.}~\bibnamefont{Vlassopoulos}},\ }%
  \bibfield{journal}{%
  \bibinfo {journal} {Phys. Rev. Lett.}\ }%
  \textbf{\bibinfo {volume} {95}},\ \bibinfo {pages} {268301} (\bibinfo {year}
  {2005})%
  \bibAnnoteFile{NoStop}{zaccarelli_tailoring_2005}%
\bibitem{stiakakis_slow_2010}%
  \BibitemOpen
  \bibfield{author}{%
  \bibinfo {author} {\bibfnamefont{E.}~\bibnamefont{Stiakakis}}, \bibinfo
  {author} {\bibfnamefont{A.}~\bibnamefont{Wilk}}, \bibinfo {author}
  {\bibfnamefont{J.}~\bibnamefont{Kohlbrecher}}, \bibinfo {author}
  {\bibfnamefont{D.}~\bibnamefont{Vlassopoulos}},\ and\ \bibinfo {author}
  {\bibfnamefont{G.}~\bibnamefont{Petekidis}},\ }%
  \bibfield{journal}{%
  \bibinfo {journal} {Phys. Rev. E}\ }%
  \textbf{\bibinfo {volume} {81}},\ \bibinfo {pages} {020402} (\bibinfo {year}
  {2010})%
  \bibAnnoteFile{NoStop}{stiakakis_slow_2010}%
\bibitem{Saunders_Vincent_1999}%
  \BibitemOpen
  \bibfield{author}{%
  \bibinfo {author} {\bibfnamefont{B.~R.}\ \bibnamefont{Saunders}}\ and\
  \bibinfo {author} {\bibfnamefont{B.}~\bibnamefont{Vincent}},\ }%
  \bibfield{journal}{%
  \bibinfo {journal} {Adv. Colloid Interface Sci.}\ }%
  \textbf{\bibinfo {volume} {80}},\ \bibinfo {pages} {1} (\bibinfo {year}
  {1999})%
  \bibAnnoteFile{NoStop}{Saunders_Vincent_1999}%
\bibitem{Senff_Richtering_1999}%
  \BibitemOpen
  \bibfield{author}{%
  \bibinfo {author} {\bibfnamefont{H.}~\bibnamefont{Senff}}\ and\ \bibinfo
  {author} {\bibfnamefont{W.}~\bibnamefont{Richtering}},\ }%
  \bibfield{journal}{%
  \bibinfo {journal} {J. Chem. Phys.}\ }%
  \textbf{\bibinfo {volume} {111}},\ \bibinfo {pages} {1705} (\bibinfo {year} {1999})%
  \bibAnnoteFile{NoStop}{Senff_Richtering_1999}%
\bibitem{Lietor-Santos_Sierra-Martin_Vavrin_Hu_Gasser_Fernandez-Nieves_2009}%
  \BibitemOpen
  \bibfield{author}{%
  \bibinfo {author} {\bibfnamefont{J.-J.}\ \bibnamefont{Lietor-Santos}},
  \bibinfo {author} {\bibfnamefont{B.}~\bibnamefont{Sierra-Martin}}, \bibinfo
  {author} {\bibfnamefont{R.}~\bibnamefont{Vavrin}}, \bibinfo {author}
  {\bibfnamefont{Z.}~\bibnamefont{Hu}}, \bibinfo {author}
  {\bibfnamefont{U.}~\bibnamefont{Gasser}},\ and\ \bibinfo {author}
  {\bibfnamefont{A.}~\bibnamefont{Fernandez-Nieves}},\ }%
  \bibfield{journal}{%
  \bibinfo {journal} {Macromolecules}\ }%
  \textbf{\bibinfo {volume} {42}},\ \bibinfo {pages} {6225} (\bibinfo {year}
  {2009})%
  \bibAnnoteFile{NoStop}{Lietor-Santos_Sierra-Martin_Vavrin_Hu_Gasser_Fernande%
z-Nieves_2009}%
\bibitem{mattsson_soft_2009}%
  \BibitemOpen
  \bibfield{author}{%
  \bibinfo {author} {\bibfnamefont{J.}~\bibnamefont{Mattsson}}, \bibinfo
  {author} {\bibfnamefont{H.~M.}\ \bibnamefont{Wyss}}, \bibinfo {author}
  {\bibfnamefont{A.}~\bibnamefont{Fernandez-Nieves}}, \bibinfo {author}
  {\bibfnamefont{K.}~\bibnamefont{Miyazaki}}, \bibinfo {author}
  {\bibfnamefont{Z.}~\bibnamefont{Hu}}, \bibinfo {author}
  {\bibfnamefont{D.~R.}\ \bibnamefont{Reichman}},\ and\ \bibinfo {author}
  {\bibfnamefont{D.~A.}\ \bibnamefont{Weitz}},\ }%
  \bibfield{journal}{%
  \bibinfo {journal} {Nature}\ }%
  \textbf{\bibinfo {volume} {462}},\ \bibinfo {pages} {83} (\bibinfo {year}
  {2009})%
  \bibAnnoteFile{NoStop}{mattsson_soft_2009}%
\bibitem{Koumakis_Petekidis_2012}%
  \BibitemOpen
  \bibfield{author}{%
  \bibinfo {author} {\bibfnamefont{N.}~\bibnamefont{Koumakis}}, \bibinfo
  {author} {\bibfnamefont{A.}~\bibnamefont{Pamvouxoglou}}, \bibinfo {author}
  {\bibfnamefont{A.~S.}\ \bibnamefont{Poulos}},\ and\ \bibinfo {author}
  {\bibfnamefont{G.}~\bibnamefont{Petekidis}},\ }%
  \bibfield{journal}{%
  \bibinfo {journal} {Soft Matter}\ }%
  \textbf{\bibinfo {volume} {8}},\ \bibinfo {pages} {4271} (\bibinfo {year}
  {2012})%
  \bibAnnoteFile{NoStop}{Koumakis_Petekidis_2012}%
\bibitem{Pamies_Cacciuto_Frenkel_2009}%
  \BibitemOpen
  \bibfield{author}{%
  \bibinfo {author} {\bibfnamefont{J.~C.}\ \bibnamefont{Pamies}}, \bibinfo
  {author} {\bibfnamefont{A.}~\bibnamefont{Cacciuto}},\ and\ \bibinfo {author}
  {\bibfnamefont{D.}~\bibnamefont{Frenkel}},\ }%
  \bibfield{journal}{%
  \bibinfo {journal} {J. Chem. Phys.}\ }%
  \textbf{\bibinfo {volume} {131}},\ \bibinfo {pages} {044514} (\bibinfo {year}
  {2009})%
  \bibAnnoteFile{NoStop}{Pamies_Cacciuto_Frenkel_2009}%
\bibitem{Stillinger1976}%
  \BibitemOpen
  \bibfield{author}{%
  \bibinfo {author} {\bibfnamefont{F.~H.}\ \bibnamefont{Stillinger}},\ }%
  \bibfield{journal}{%
  \bibinfo {journal} {J. Chem. Phys.}\ }%
  \textbf{\bibinfo {volume} {65}},\ \bibinfo {pages} {3968} (\bibinfo {year}
  {1976})%
  \bibAnnoteFile{NoStop}{Stillinger1976}%
\bibitem{Stillinger1979}%
  \BibitemOpen
  \bibfield{author}{%
  \bibinfo {author} {\bibfnamefont{F.~H.}\ \bibnamefont{Stillinger}},\ }%
  \bibfield{journal}{%
  \bibinfo {journal} {Phys. Rev. B}\ }%
  \textbf{\bibinfo {volume} {20}},\ \bibinfo {pages} {299} (\bibinfo {year}
  {1979})%
  \bibAnnoteFile{NoStop}{Stillinger1979}%
\bibitem{louis_mean-field_2000}%
  \BibitemOpen
  \bibfield{author}{%
  \bibinfo {author} {\bibfnamefont{A.~A.}\ \bibnamefont{Louis}}, \bibinfo
  {author} {\bibfnamefont{P.~G.}\ \bibnamefont{Bolhuis}},\ and\ \bibinfo
  {author} {\bibfnamefont{J.~P.}\ \bibnamefont{Hansen}},\ }%
  \bibfield{journal}{%
  \bibinfo {journal} {Phys. Rev. E}\ }%
  \textbf{\bibinfo {volume} {62}},\ \bibinfo {pages} {7961} (\bibinfo {year}
  {2000})%
  \bibAnnoteFile{NoStop}{louis_mean-field_2000}%
\bibitem{lang_fluid_2000}%
  \BibitemOpen
  \bibfield{author}{%
  \bibinfo {author} {\bibfnamefont{A.}~\bibnamefont{Lang}}, \bibinfo {author}
  {\bibfnamefont{C.~N.}\ \bibnamefont{Likos}}, \bibinfo {author}
  {\bibfnamefont{M.}~\bibnamefont{Watzlawek}},\ and\ \bibinfo {author}
  {\bibfnamefont{H.}~\bibnamefont{Lowen}},\ }%
  \bibfield{journal}{%
  \bibinfo {journal} {J. Phys.: Condens. Matter}\ }%
  \textbf{\bibinfo {volume} {12}},\ \bibinfo {pages} {5087} (\bibinfo {year}
  {2000})%
  \bibAnnoteFile{NoStop}{lang_fluid_2000}%
\bibitem{prestipino_phase_2005}%
  \BibitemOpen
  \bibfield{author}{%
  \bibinfo {author} {\bibfnamefont{S.}~\bibnamefont{Prestipino}}, \bibinfo
  {author} {\bibfnamefont{F.}~\bibnamefont{Saija}},\ and\ \bibinfo {author}
  {\bibfnamefont{P.~V.}\ \bibnamefont{Giaquinta}},\ }%
  \bibfield{journal}{%
  \bibinfo {journal} {Phys. Rev. E}\ }%
  \textbf{\bibinfo {volume} {71}},\ \bibinfo {pages} {050102} (\bibinfo {year}
  {2005})%
  \bibAnnoteFile{NoStop}{prestipino_phase_2005}%
\bibitem{Mausbach2006}%
  \BibitemOpen
  \bibfield{author}{%
  \bibinfo {author} {\bibfnamefont{P.}~\bibnamefont{Mausbach}}\ and\ \bibinfo
  {author} {\bibfnamefont{H.-O.}\ \bibnamefont{May}},\ }%
  \bibfield{journal}{%
  \bibinfo {journal} {Fluid Phase Equilibria}\ }%
  \textbf{\bibinfo {volume} {249}},\ \bibinfo {pages} {17} (\bibinfo {year}
  {2006})%
  \bibAnnoteFile{NoStop}{Mausbach2006}%
\bibitem{mayer_asymmetric_2008}%
  \BibitemOpen
  \bibfield{author}{%
  \bibinfo {author} {\bibfnamefont{C.}~\bibnamefont{Mayer}}, \bibinfo {author}
  {\bibfnamefont{E.}~\bibnamefont{Zaccarelli}}, \bibinfo {author}
  {\bibfnamefont{E.}~\bibnamefont{Stiakakis}}, \bibinfo {author}
  {\bibfnamefont{C.~N.}\ \bibnamefont{Likos}}, \bibinfo {author}
  {\bibfnamefont{F.}~\bibnamefont{Sciortino}}, \bibinfo {author}
  {\bibfnamefont{A.}~\bibnamefont{Munam}}, \bibinfo {author}
  {\bibfnamefont{M.}~\bibnamefont{Gauthier}}, \bibinfo {author}
  {\bibfnamefont{N.}~\bibnamefont{Hadjichristidis}}, \bibinfo {author}
  {\bibfnamefont{H.}~\bibnamefont{Iatrou}}, \bibinfo {author}
  {\bibfnamefont{P.}~\bibnamefont{Tartaglia}}, \bibinfo {author}
  {\bibfnamefont{H.}~\bibnamefont{Lowen}},\ and\ \bibinfo {author}
  {\bibfnamefont{D.}~\bibnamefont{Vlassopoulos}},\ }%
  \bibfield{journal}{%
  \bibinfo {journal} {Nature Mat.}\ }%
  \textbf{\bibinfo {volume} {7}},\ \bibinfo {pages} {780} (\bibinfo {year}
  {2008})%
  \bibAnnoteFile{NoStop}{mayer_asymmetric_2008}%
\bibitem{krekelberg_anomalous_2009}%
  \BibitemOpen
  \bibfield{author}{%
  \bibinfo {author} {\bibfnamefont{W.~P.}\ \bibnamefont{Krekelberg}}, \bibinfo
  {author} {\bibfnamefont{T.}~\bibnamefont{Kumar}}, \bibinfo {author}
  {\bibfnamefont{J.}~\bibnamefont{Mittal}}, \bibinfo {author}
  {\bibfnamefont{J.~R.}\ \bibnamefont{Errington}},\ and\ \bibinfo {author}
  {\bibfnamefont{T.~M.}\ \bibnamefont{Truskett}},\ }%
  \bibfield{journal}{%
  \bibinfo {journal} {Phys. Rev. E}\ }%
  \textbf{\bibinfo {volume} {79}},\ \bibinfo {pages} {031203} (\bibinfo {year}
  {2009})%
  \bibAnnoteFile{NoStop}{krekelberg_anomalous_2009}%
\bibitem{krekelberg_generalized_2009}%
  \BibitemOpen
  \bibfield{author}{%
  \bibinfo {author} {\bibfnamefont{W.}~\bibnamefont{Krekelberg}}, \bibinfo
  {author} {\bibfnamefont{M.}~\bibnamefont{Pond}}, \bibinfo {author}
  {\bibfnamefont{G.}~\bibnamefont{Goel}}, \bibinfo {author}
  {\bibfnamefont{V.}~\bibnamefont{Shen}}, \bibinfo {author}
  {\bibfnamefont{J.}~\bibnamefont{Errington}},\ and\ \bibinfo {author}
  {\bibfnamefont{T.}~\bibnamefont{Truskett}},\ }%
  \bibfield{journal}{%
  \bibinfo {journal} {Phys. Rev. E}\ }%
  \textbf{\bibinfo {volume} {80}},\ \bibinfo {pages} {061205} (\bibinfo {year}
  {2009})%
  \bibAnnoteFile{NoStop}{krekelberg_generalized_2009}%
\bibitem{Jacquin_Berthier_2010}%
  \BibitemOpen
  \bibfield{author}{%
  \bibinfo {author} {\bibfnamefont{H.}~\bibnamefont{Jacquin}}\ and\ \bibinfo
  {author} {\bibfnamefont{L.}~\bibnamefont{Berthier}},\ }%
  \bibfield{journal}{%
  \bibinfo {journal} {Soft Matter}\ }%
  \textbf{\bibinfo {volume} {6}},\ \bibinfo {pages} {2970} (\bibinfo {month}
  {Jun}\ \bibinfo {year} {2010})%
  \bibAnnoteFile{NoStop}{Jacquin_Berthier_2010}%
\bibitem{berthier_increasing_2010}%
  \BibitemOpen
  \bibfield{author}{%
  \bibinfo {author} {\bibfnamefont{L.}~\bibnamefont{Berthier}}, \bibinfo
  {author} {\bibfnamefont{A.~J.}\ \bibnamefont{Moreno}},\ and\ \bibinfo
  {author} {\bibfnamefont{G.}~\bibnamefont{Szamel}},\ }%
  \bibfield{journal}{%
  \bibinfo {journal} {Phys. Rev. E}\ }%
  \textbf{\bibinfo {volume} {82}},\ \bibinfo {pages} {060501} (\bibinfo {year}
  {2010})%
  \bibAnnoteFile{NoStop}{berthier_increasing_2010}%
\bibitem{ikeda_glass_2011}%
  \BibitemOpen
  \bibfield{author}{%
  \bibinfo {author} {\bibfnamefont{A.}~\bibnamefont{Ikeda}}\ and\ \bibinfo
  {author} {\bibfnamefont{K.}~\bibnamefont{Miyazaki}},\ }%
  \bibfield{journal}{%
  \bibinfo {journal} {Phys. Rev. Lett.}\ }%
  \textbf{\bibinfo {volume} {106}},\ \bibinfo {pages} {015701} (\bibinfo {year}
  {2011})%
  \bibAnnoteFile{NoStop}{ikeda_glass_2011}%
\bibitem{Ikeda_therm_2011}%
  \BibitemOpen
  \bibfield{author}{%
  \bibinfo {author} {\bibfnamefont{A.}~\bibnamefont{Ikeda}}\ and\ \bibinfo
  {author} {\bibfnamefont{K.}~\bibnamefont{Miyazaki}},\ }%
  \bibfield{journal}{%
  \bibinfo {journal} {J. Chem. Phys.}\ }%
  \textbf{\bibinfo {volume} {135}},\ \bibinfo {pages} {024901} (\bibinfo {year}
  {2011})%
  \bibAnnoteFile{NoStop}{Ikeda_therm_2011}%
\bibitem{ikeda_slow_2011}%
  \BibitemOpen
  \bibfield{author}{%
  \bibinfo {author} {\bibfnamefont{A.}~\bibnamefont{Ikeda}}\ and\ \bibinfo
  {author} {\bibfnamefont{K.}~\bibnamefont{Miyazaki}},\ }%
  \bibfield{journal}{%
  \bibinfo {journal} {J. Chem. Phys.}\ }%
  \textbf{\bibinfo {volume} {135}},\ \bibinfo {pages} {054901} (\bibinfo {year}
  {2011})%
  \bibAnnoteFile{NoStop}{ikeda_slow_2011}%
\bibitem{Paloli_Mohanty_Crassous_Zaccarelli_Schurtenberger_2013}%
  \BibitemOpen
  \bibfield{author}{%
  \bibinfo {author} {\bibfnamefont{D.}~\bibnamefont{Paloli}}, \bibinfo {author}
  {\bibfnamefont{P.~S.}\ \bibnamefont{Mohanty}}, \bibinfo {author}
  {\bibfnamefont{J.~J.}\ \bibnamefont{Crassous}}, \bibinfo {author}
  {\bibfnamefont{E.}~\bibnamefont{Zaccarelli}},\ and\ \bibinfo {author}
  {\bibfnamefont{P.}~\bibnamefont{Schurtenberger}},\ }%
  \bibfield{journal}{%
  \bibinfo {journal} {Soft Matter}\ }%
  \textbf{\bibinfo {volume} {9}},\ \bibinfo {pages} {3000} (\bibinfo {year}
  {2013})%
  \bibAnnoteFile{NoStop}{Paloli_Mohanty_Crassous_Zaccarelli_Schurtenberger_201%
3}%
\bibitem{Marquest_1989}%
  \BibitemOpen
  \bibfield{author}{%
  \bibinfo {author} {\bibfnamefont{C.}~\bibnamefont{Marquest}}\ and\ \bibinfo
  {author} {\bibfnamefont{T.~A.}\ \bibnamefont{Witten}},\ }%
  \bibfield{journal}{%
  \Doi{10.1051/jphys:0198900500100126700}{\bibinfo {journal} {J. Phys.
  France}}\ }%
  \textbf{\bibinfo {volume} {50}},\ \bibinfo {pages} {1267} (\bibinfo {year}
  {1989})%
  \bibAnnoteFile{NoStop}{Marquest_1989}%
\bibitem{Likos_PSM_1998}%
  \BibitemOpen
  \bibfield{author}{%
  \bibinfo {author} {\bibfnamefont{C.~N.}\ \bibnamefont{Likos}}, \bibinfo
  {author} {\bibfnamefont{M.}~\bibnamefont{Watzlawek}},\ and\ \bibinfo {author}
  {\bibfnamefont{H.}~\bibnamefont{L\"owen}},\ }%
  \bibfield{journal}{%
  \bibinfo {journal} {Phys. Rev. E}\ }%
  \textbf{\bibinfo {volume} {58}},\ \bibinfo {pages} {3135} (\bibinfo {year}
  {1998})%
  \bibAnnoteFile{NoStop}{Likos_PSM_1998}%
\bibitem{mladek_formation_2006}%
  \BibitemOpen
  \bibfield{author}{%
  \bibinfo {author} {\bibfnamefont{B.~M.}\ \bibnamefont{Mladek}}, \bibinfo
  {author} {\bibfnamefont{D.}~\bibnamefont{Gottwald}}, \bibinfo {author}
  {\bibfnamefont{G.}~\bibnamefont{Kahl}}, \bibinfo {author}
  {\bibfnamefont{M.}~\bibnamefont{Neumann}},\ and\ \bibinfo {author}
  {\bibfnamefont{C.~N.}\ \bibnamefont{Likos}},\ }%
  \bibfield{journal}{%
  \bibinfo {journal} {Phys. Rev. Lett.}\ }%
  \textbf{\bibinfo {volume} {96}},\ \bibinfo {pages} {045701} (\bibinfo {year}
  {2006})%
  \bibAnnoteFile{NoStop}{mladek_formation_2006}%
\bibitem{mladek_clustering_2007}%
  \BibitemOpen
  \bibfield{author}{%
  \bibinfo {author} {\bibfnamefont{B.~M.}\ \bibnamefont{Mladek}}, \bibinfo
  {author} {\bibfnamefont{D.}~\bibnamefont{Gottwald}}, \bibinfo {author}
  {\bibfnamefont{G.}~\bibnamefont{Kahl}}, \bibinfo {author}
  {\bibfnamefont{M.}~\bibnamefont{Neumann}},\ and\ \bibinfo {author}
  {\bibfnamefont{C.~N.}\ \bibnamefont{Likos}},\ }%
  \bibfield{journal}{%
  \bibinfo {journal} {J. Phys. Chem. B}\ }%
  \textbf{\bibinfo {volume} {111}},\ \bibinfo {pages} {12799} (\bibinfo {year}
  {2007})%
  \bibAnnoteFile{NoStop}{mladek_clustering_2007}%
\bibitem{Zhang_Charbonneau_Mladek_2010}%
  \BibitemOpen
  \bibfield{author}{%
  \bibinfo {author} {\bibfnamefont{K.}~\bibnamefont{Zhang}}, \bibinfo {author}
  {\bibfnamefont{P.}~\bibnamefont{Charbonneau}},\ and\ \bibinfo {author}
  {\bibfnamefont{B.~M.}\ \bibnamefont{Mladek}},\ }%
  \bibfield{journal}{%
  \bibinfo {journal} {Phys. Rev. Lett.}\ }%
  \textbf{\bibinfo {volume} {105}},\ \bibinfo {pages} {245701} (\bibinfo {year}
  {2010})%
  \bibAnnoteFile{NoStop}{Zhang_Charbonneau_Mladek_2010}%
\bibitem{Zhang_PSM_2012}%
  \BibitemOpen
  \bibfield{author}{%
  \bibinfo {author} {\bibfnamefont{K.}~\bibnamefont{Zhang}}\ and\ \bibinfo
  {author} {\bibfnamefont{P.}~\bibnamefont{Charbonneau}},\ }%
  \bibfield{journal}{%
  \bibinfo {journal} {J. Chem. Phys.}\ }%
  \textbf{\bibinfo {volume} {136}},\ \bibinfo {eid} {214106} (\bibinfo {year}
  {2012})%
  \bibAnnoteFile{NoStop}{Zhang_PSM_2012}%
\bibitem{coslovich_cluster_2012}%
  \BibitemOpen
  \bibfield{author}{%
  \bibinfo {author} {\bibfnamefont{D.}~\bibnamefont{Coslovich}}, \bibinfo
  {author} {\bibfnamefont{M.}~\bibnamefont{Bernabei}},\ and\ \bibinfo {author}
  {\bibfnamefont{A.~J.}\ \bibnamefont{Moreno}},\ }%
  \bibfield{journal}{%
  \bibinfo {journal} {J. Chem. Phys.}\ }%
  \textbf{\bibinfo {volume} {137}},\ \bibinfo {eid} {184904} (\bibinfo {year}
  {2012})%
  \bibAnnoteFile{NoStop}{coslovich_cluster_2012}%
\bibitem{moreno_diffusion_2007}%
  \BibitemOpen
  \bibfield{author}{%
  \bibinfo {author} {\bibfnamefont{A.~J.}\ \bibnamefont{Moreno}}\ and\ \bibinfo
  {author} {\bibfnamefont{C.~N.}\ \bibnamefont{Likos}},\ }%
  \bibfield{journal}{%
  \bibinfo {journal} {Phys. Rev. Lett.}\ }%
  \textbf{\bibinfo {volume} {99}},\ \bibinfo {pages} {107801} (\bibinfo {year}
  {2007})%
  \bibAnnoteFile{NoStop}{moreno_diffusion_2007}%
\bibitem{coslovich_hopping_2011}%
  \BibitemOpen
  \bibfield{author}{%
  \bibinfo {author} {\bibfnamefont{D.}~\bibnamefont{Coslovich}}, \bibinfo
  {author} {\bibfnamefont{L.}~\bibnamefont{Strauss}},\ and\ \bibinfo {author}
  {\bibfnamefont{G.}~\bibnamefont{Kahl}},\ }%
  \bibfield{journal}{%
  \bibinfo {journal} {Soft Matter}\ }%
  \textbf{\bibinfo {volume} {7}},\ \bibinfo {pages} {2127} (\bibinfo {year}
  {2011})%
  \bibAnnoteFile{NoStop}{coslovich_hopping_2011}%
\bibitem{Likos_Lang_Watzlawek_Lowen_2001}%
  \BibitemOpen
  \bibfield{author}{%
  \bibinfo {author} {\bibfnamefont{C.~N.}\ \bibnamefont{Likos}}, \bibinfo
  {author} {\bibfnamefont{A.}~\bibnamefont{Lang}}, \bibinfo {author}
  {\bibfnamefont{M.}~\bibnamefont{Watzlawek}},\ and\ \bibinfo {author}
  {\bibfnamefont{H.}~\bibnamefont{L{\"o}wen}},\ }%
  \bibfield{journal}{%
  \bibinfo {journal} {Phys. Rev. E}\ }%
  \textbf{\bibinfo {volume} {63}},\ \bibinfo {pages} {031206} (\bibinfo {year}
  {2001})%
  \bibAnnoteFile{NoStop}{Likos_Lang_Watzlawek_Lowen_2001}%
\bibitem{likos_why_2007}%
  \BibitemOpen
  \bibfield{author}{%
  \bibinfo {author} {\bibfnamefont{C.~N.}\ \bibnamefont{Likos}}, \bibinfo
  {author} {\bibfnamefont{B.~M.}\ \bibnamefont{Mladek}}, \bibinfo {author}
  {\bibfnamefont{D.}~\bibnamefont{Gottwald}},\ and\ \bibinfo {author}
  {\bibfnamefont{G.}~\bibnamefont{Kahl}},\ }%
  \bibfield{journal}{%
  \bibinfo {journal} {J. Chem. Phys.}\ }%
  \textbf{\bibinfo {volume} {126}},\ \bibinfo {pages} {224502} (\bibinfo {year}
  {2007})%
  \bibAnnoteFile{NoStop}{likos_why_2007}%
\bibitem{mladek_computer_2008}%
  \BibitemOpen
  \bibfield{author}{%
  \bibinfo {author} {\bibfnamefont{B.~M.}\ \bibnamefont{Mladek}}, \bibinfo
  {author} {\bibfnamefont{G.}~\bibnamefont{Kahl}},\ and\ \bibinfo {author}
  {\bibfnamefont{C.~N.}\ \bibnamefont{Likos}},\ }%
  \bibfield{journal}{%
  \bibinfo {journal} {Phys. Rev. Lett.}\ }%
  \textbf{\bibinfo {volume} {100}},\ \bibinfo {pages} {028301} (\bibinfo {year}
  {2008})%
  \bibAnnoteFile{NoStop}{mladek_computer_2008}%
\bibitem{Lenz_Cluster_2012}%
  \BibitemOpen
  \bibfield{author}{%
  \bibinfo {author} {\bibfnamefont{D.~A.}\ \bibnamefont{Lenz}}, \bibinfo
  {author} {\bibfnamefont{R.}~\bibnamefont{Blaak}}, \bibinfo {author}
  {\bibfnamefont{C.~N.}\ \bibnamefont{Likos}},\ and\ \bibinfo {author}
  {\bibfnamefont{B.~M.}\ \bibnamefont{Mladek}},\ }%
  \bibfield{journal}{%
  \bibinfo {journal} {Phys. Rev. Lett.}\ }%
  \textbf{\bibinfo {volume} {109}},\ \bibinfo {pages} {228301} (\bibinfo {year}
  {2012})%
  \bibAnnoteFile{NoStop}{Lenz_Cluster_2012}%
\bibitem{sastry__1998}%
  \BibitemOpen
  \bibfield{author}{%
  \bibinfo {author} {\bibfnamefont{S.}~\bibnamefont{Sastry}}, \bibinfo {author}
  {\bibfnamefont{P.~G.}\ \bibnamefont{Debenedetti}},\ and\ \bibinfo {author}
  {\bibfnamefont{F.~H.}\ \bibnamefont{Stillinger}},\ }%
  \bibfield{journal}{%
  \bibinfo {journal} {Nature}\ }%
  \textbf{\bibinfo {volume} {393}},\ \bibinfo {pages} {554} (\bibinfo {year}
  {1998})%
  \bibAnnoteFile{NoStop}{sastry__1998}%
\bibitem{hansen_theory_2006-1}%
  \BibitemOpen
  \bibfield{author}{%
  \bibinfo {author} {\bibfnamefont{J.-P.}\ \bibnamefont{Hansen}}\ and\ \bibinfo
  {author} {\bibfnamefont{I.~R.}\ \bibnamefont{{McDonald}}},\ }%
  \emph{\bibinfo {title} {Theory of Simple Liquids}},\ \bibinfo {edition}
  {3rd}\ ed.\ (\bibinfo {publisher} {Academic Press},\ \bibinfo {address} {New
  York},\ \bibinfo {year} {2006})%
  \bibAnnoteFile{NoStop}{hansen_theory_2006-1}%
\bibitem{ingebrigtsen_what_2012}%
  \BibitemOpen
  \bibfield{author}{%
  \bibinfo {author} {\bibfnamefont{T.~S.}\ \bibnamefont{Ingebrigtsen}},
  \bibinfo {author} {\bibfnamefont{T.~B.}\ \bibnamefont{Schr{\o}der}},\ and\
  \bibinfo {author} {\bibfnamefont{J.~C.}\ \bibnamefont{Dyre}},\ }%
  \bibfield{journal}{%
  \bibinfo {journal} {Phys. Rev. X}\ }%
  \textbf{\bibinfo {volume} {2}},\ \bibinfo {pages} {011011} (\bibinfo {year}
  {2012})%
  \bibAnnoteFile{NoStop}{ingebrigtsen_what_2012}%
\bibitem{grigera_geometric_2002-1}%
  \BibitemOpen
  \bibfield{author}{%
  \bibinfo {author} {\bibfnamefont{T.~S.}\ \bibnamefont{Grigera}}, \bibinfo
  {author} {\bibfnamefont{A.}~\bibnamefont{Cavagna}}, \bibinfo {author}
  {\bibfnamefont{I.}~\bibnamefont{Giardina}},\ and\ \bibinfo {author}
  {\bibfnamefont{G.}~\bibnamefont{Parisi}},\ }%
  \bibfield{journal}{%
  \bibinfo {journal} {Phys. Rev. Lett.}\ }%
  \textbf{\bibinfo {volume} {88}},\ \bibinfo {pages} {055502} (\bibinfo {year}
  {2002})%
  \bibAnnoteFile{NoStop}{grigera_geometric_2002-1}%
\bibitem{stillinger__1982}%
  \BibitemOpen
  \bibfield{author}{%
  \bibinfo {author} {\bibfnamefont{F.~H.}\ \bibnamefont{Stillinger}}\ and\
  \bibinfo {author} {\bibfnamefont{T.~A.}\ \bibnamefont{Weber}},\ }%
  \bibfield{journal}{%
  \bibinfo {journal} {Phys. Rev A}\ }%
  \textbf{\bibinfo {volume} {25}},\ \bibinfo {pages} {983} (\bibinfo {year}
  {1982})%
  \bibAnnoteFile{NoStop}{stillinger__1982}%
\bibitem{liu__1989}%
  \BibitemOpen
  \bibfield{author}{%
  \bibinfo {author} {\bibfnamefont{D.~C.}\ \bibnamefont{Liu}}\ and\ \bibinfo
  {author} {\bibfnamefont{J.}~\bibnamefont{Nocedal}},\ }%
  \bibfield{journal}{%
  \bibinfo {journal} {Math. Program.}\ }%
  \textbf{\bibinfo {volume} {45}},\ \bibinfo {pages} {503} (\bibinfo {year}
  {1989})%
  \bibAnnoteFile{NoStop}{liu__1989}%
\bibitem{likos_effective_2001}%
  \BibitemOpen
  \bibfield{author}{%
  \bibinfo {author} {\bibfnamefont{C.~N.}\ \bibnamefont{Likos}},\ }%
  \bibfield{journal}{%
  \bibinfo {journal} {Phys. Rep.}\ }%
  \textbf{\bibinfo {volume} {348}},\ \bibinfo {pages} {267} (\bibinfo {year}
  {2001})%
  \bibAnnoteFile{NoStop}{likos_effective_2001}%
\bibitem{Hansen_Verlet_1969}%
  \BibitemOpen
  \bibfield{author}{%
  \bibinfo {author} {\bibfnamefont{J.-P.}\ \bibnamefont{Hansen}}\ and\ \bibinfo
  {author} {\bibfnamefont{L.}~\bibnamefont{Verlet}},\ }%
  \bibfield{journal}{%
  \bibinfo {journal} {Phys. Rev.}\ }%
  \textbf{\bibinfo {volume} {184}},\ \bibinfo {pages} {151} (\bibinfo {year}
  {1969})%
  \bibAnnoteFile{NoStop}{Hansen_Verlet_1969}%
\bibitem{Neuhaus_Likos_2011}%
  \BibitemOpen
  \bibfield{author}{%
  \bibinfo {author} {\bibfnamefont{T.}~\bibnamefont{Neuhaus}}\ and\ \bibinfo
  {author} {\bibfnamefont{C.~N.}\ \bibnamefont{Likos}},\ }%
  \bibfield{journal}{%
  \bibinfo {journal} {J. Phys.: Condens. Matter}\ }%
  \textbf{\bibinfo {volume} {23}},\ \bibinfo {pages} {234112} (\bibinfo {year}
  {2011})%
  \bibAnnoteFile{NoStop}{Neuhaus_Likos_2011}%
\bibitem{sarkar_inherent_2011}%
  \BibitemOpen
  \bibfield{author}{%
  \bibinfo {author} {\bibfnamefont{S.}~\bibnamefont{Sarkar}}\ and\ \bibinfo
  {author} {\bibfnamefont{B.}~\bibnamefont{Bagchi}},\ }%
  \bibfield{journal}{%
  \bibinfo {journal} {Phys. Rev. E}\ }%
  \textbf{\bibinfo {volume} {83}},\ \bibinfo {pages} {031506} (\bibinfo {year}
  {2011})%
  \bibAnnoteFile{NoStop}{sarkar_inherent_2011}%
\bibitem{Fomin_Ryzhov_Gribova_2010}%
  \BibitemOpen
  \bibfield{author}{%
  \bibinfo {author} {\bibfnamefont{Y.~D.}\ \bibnamefont{Fomin}}, \bibinfo
  {author} {\bibfnamefont{V.~N.}\ \bibnamefont{Ryzhov}},\ and\ \bibinfo
  {author} {\bibfnamefont{N.~V.}\ \bibnamefont{Gribova}},\ }%
  \bibfield{journal}{%
  \bibinfo {journal} {Phys. Rev. E}\ }%
  \textbf{\bibinfo {volume} {81}},\ \bibinfo {pages} {061201} (\bibinfo {year}
  {2010})%
  \bibAnnoteFile{NoStop}{Fomin_Ryzhov_Gribova_2010}%
\bibitem{rosenfeld_quasi-universal_1999}%
  \BibitemOpen
  \bibfield{author}{%
  \bibinfo {author} {\bibfnamefont{Y.}~\bibnamefont{Rosenfeld}},\ }%
  \bibfield{journal}{%
  \bibinfo {journal} {J. Phys.: Condens. Matter}\ }%
  \textbf{\bibinfo {volume} {11}},\ \bibinfo {pages} {5415} (\bibinfo {year}
  {1999})%
  \bibAnnoteFile{NoStop}{rosenfeld_quasi-universal_1999}%
\bibitem{Pond_Errington_Truskett_2011}%
  \BibitemOpen
  \bibfield{author}{%
  \bibinfo {author} {\bibfnamefont{M.~J.}\ \bibnamefont{Pond}}, \bibinfo
  {author} {\bibfnamefont{J.~R.}\ \bibnamefont{Errington}},\ and\ \bibinfo
  {author} {\bibfnamefont{T.~M.}\ \bibnamefont{Truskett}},\ }%
  \bibfield{journal}{%
  \bibinfo {journal} {Soft Matter}\ }%
  \textbf{\bibinfo {volume} {7}},\ \bibinfo {pages} {9859} (\bibinfo {year}
  {2011})%
  \bibAnnoteFile{NoStop}{Pond_Errington_Truskett_2011}%
\end{thebibliography}

%

\end{document}